\documentclass[a4paper,11pt]{article}
\usepackage[margin=1in]{geometry}
\usepackage{mathpazo}
\usepackage{microtype}
\makeatletter
\renewcommand\paragraph{\@startsection{paragraph}{4}{\z@}%
  {1.2ex \@plus .4ex \@minus .2ex}{-1em}{\normalfont\normalsize\bfseries}}
\makeatother
\usepackage{amsmath,amssymb,amsthm}
\usepackage{graphicx}
\usepackage{booktabs}
\usepackage{array}
\usepackage[hyphens]{url}
\usepackage{hyperref}
\hypersetup{colorlinks=true,linkcolor=blue!50!black,citecolor=green!40!black,urlcolor=blue!60!black}
\usepackage{xcolor}
\usepackage{enumitem}
\usepackage{placeins}
\usepackage{tikz}
\usepackage[normalem]{ulem}
\usetikzlibrary{arrows.meta,positioning}

\newcommand{\num}[1]{#1}
\newcommand{\wprov}{WP}

\newcommand{\link}{\mathsf{BA}}
\newcommand{\linker}{\mathsf{BAI}}
\newcommand{\poolattack}{Credential Pooling Attack}
\newcommand{\proto}{\textsf{CRESTA}}
\newif\ifcomments
\commentstrue
\definecolor{ivancolor}{RGB}{0,90,170}

\newcommand{\ignore}[1]{}

\title{\Large\bfseries Fooling Relying Parties of the EUDI Wallet Architecture:\\
Credential Pooling Attacks and\\ \proto, a Standard-Compliant Countermeasure}

\author{
\begin{tabular}[t]{c}
Leonardo Rufini\\
\normalsize\texttt{leonardo.rufini@uniroma1.it}
\end{tabular}
\qquad
\begin{tabular}[t]{c}
Ivan Visconti\\
\normalsize\texttt{visconti@diag.uniroma1.it}
\end{tabular}\\[1.0em]
\normalsize DIAG, Sapienza University of Rome, Italy
}

\date{}

\begin{document}
\maketitle

\begin{abstract}

The European Digital Identity (EUDI) Wallet will let millions of EU citizens prove who they are and what they are entitled to, disclosing only what is needed. Many services need several credentials at once, such as an identity card and a university degree, so combined presentations are essential.

In this work we study credential pooling attacks on relying parties of the latest version of the EUDI Wallet reference architecture. Colluding users, each holding some credentials in their own wallet, cooperate through a proxy that acts as a single synthetic user and presents all the requested credentials, although no single user holds them all. Every signature is genuine and computed by certified hardware. We tested the attack against the European Commission's reference implementation, using two Android smartphones.

The attack would fail if all credentials of a user carried the same public key, but, in such a case, colluding relying parties could then link all presentations of that user. For this reason, the Architecture and Reference Framework (ARF) requires a fresh key for each credential and recommends batches of single-use copies, each with its own key. The use of unrelated keys is precisely what makes the pooling attack possible. 
One might think that preventing pooling requires giving up unlinkability or efficiency, or adopting non-standard cryptography, such as zero-knowledge proofs.

We show, perhaps surprisingly, that this is not the case: we present CRESTA (CREdential STApling), an efficient countermeasure that uses only standard cryptography and credential formats and preserves unlinkability. In CRESTA, an issuer checks at issuance that some keys were generated in the same secure hardware, and certifies it in a binding attestation: a standard credential, released in a batch of copies, whose attributes can be selectively disclosed. The user will decide which linked keys to disclose at presentation time. CRESTA works entirely and efficiently within the EUDI Wallet framework.
\end{abstract}

\section{Introduction}
\label{sec:intro}

The eIDAS~2.0 regulation~\cite{eidas2}, the EU legal framework for
electronic identification, requires every EU Member State to offer its
citizens a European Digital Identity (EUDI) Wallet by the end of 2026.
The wallet is specified in the Architecture and Reference Framework
(ARF)~\cite{arf}, a technical document published by the European
Commission together with Member State experts.

The wallet is a smartphone application that stores credentials:
sets of attributes about a person, signed by the organization that vouches
for them, the issuer. The most important credential is the Person
Identification Data (PID), the digital counterpart of a national
identity card. All other credentials, such as diplomas or driving
licenses, are called Electronic Attestations of Attributes
(EAAs). A service that needs to learn something about the user,
called a relying party, receives a presentation of the
requested credentials, in which only the requested attributes are
revealed. Credentials are encoded as ISO~18013-5 mdocs~\cite{iso18013} or
as SD-JWT verifiable credentials (SD-JWT VCs)~\cite{sdjwtvc}, are signed
by issuers with ECDSA, and are exchanged through OpenID for Verifiable
Credential Issuance (OpenID4VCI)~\cite{oid4vci} and OpenID for Verifiable
Presentations (OpenID4VP)~\cite{oid4vp}. Table~\ref{tab:terms} in
Section~\ref{sec:part:background} summarizes the terms used in this paper.

\paragraph{Device binding.}
At issuance, the wallet
generates a key pair inside its Wallet Secure Cryptographic Device
(WSCD), a certified tamper-resistant component of the smartphone from
which the private key cannot be extracted. A Key Attestation (KA), signed by the Wallet Provider, an organization mandated or recognized by a Member State that provides the wallet app and establishes the trust mechanisms for hardware security,
tells the issuer that the key was generated in a genuine WSCD, and the
issuer includes the public key in the credential. When presenting the
credential, the wallet signs with
that key a fresh challenge sent by the relying party, which proves that it holds the private key.
This mechanism prevents the cloning of a credential, which therefore guarantees that a credential can only be used with the active online participation of the smartphone that generated the corresponding public key. 

\paragraph{Relying-party unlinkability and batch issuance.}
The ARF also provides a mechanism for relying-party unlinkability, that is, it prevents colluding relying parties from telling whether two presentations come
from the same user. 
This feature of the ARF
relies on the batch issuance of a credential, which gives the wallet several single-use copies, each bound to a fresh key, with the wallet using
one copy per presentation.

\paragraph{Pooling resistance.}
Many real-world services require several credentials at once: a car
rental company asks for a PID and a driving license, an employer for a PID
and a university degree. The relying party then needs to know that all presented credentials belong to the same person, a property we call pooling resistance.
\label{sec:related}
In~\cite{cryptofeedback}, it is remarked  within a broader
privacy argument that batch issuance  may make credential sharing hard to prevent.
Since the keys of a user look uncorrelated, nothing in a presentation
shows that the presented credentials belong to the same user. 
In this respect, the ARF 
sets high-level requirements for a \emph{cryptographic binding of
attestations}, envisioning an issuer that verifies a proof that a new key was generated in the same WSCD as an existing one~\cite{arf_annex2}.
However, the ARF does not
specify any explicit mechanism that realizes this binding, leaving open the question of simultaneously avoiding pooling while maintaining unlinkability and usability.
The discussion topics associated with the ARF mention the ``risk of malicious combinations''~\cite{arf_topic_k}, and the problem had also been raised in the W3C credentials community~\cite{w3c1136}.
The risks of such attacks, initially envisioned in~\cite{cryptofeedback}, were also more concretely  illustrated by Verheul in~\cite{verheul} where in particular the need is remarked that a relying party be able to verify that the keys of different attestations are bound to the same WSCD. Verheul also proposed a countermeasure \cite[Section~2.1, Annex~D]{verheul}, but it requires the WSCD to perform computations on private keys which currently deployed secure hardware does not allow~\cite{arf_disc442}, and relying parties to verify an additional zero-knowledge proof, so it is out of the bounds of the ARF. Our countermeasure instead uses only operations available in a large number of currently available smartphones  (Section~\ref{sec:se}). 

\paragraph{Staying within the bounds of the ARF.}

The main challenge is to defeat pooling attacks while maintaining usability and unlinkability and remaining within the bounds of the ARF. 
More concretely, we identify the following requirements:
\begin{enumerate}[label=C\arabic*,leftmargin=2.2em,itemsep=1pt,topsep=2pt]
\item\label{c:se} \textbf{WSCD unchanged.} The WSCD is used only through
operations largely available in the secure hardware of existing smartphones, such as generating keys,
attesting them and computing ECDSA signatures (Section~\ref{sec:se});

\item\label{c:fmt} \textbf{Credentials unchanged.} Issuers keep producing
ECDSA-signed mdocs and SD-JWT VCs, each with a fresh device key as the ARF requires~\cite[ISSU\_12b]{arf_annex2};
\item\label{c:ver} \textbf{Relying parties unchanged in kind.} Relying
parties are only involved in presentations of standard credentials to check attributes according to their known access control policies;
\item\label{c:unlink} \textbf{Relying-party unlinkability.} Colluding
relying parties cannot tell whether two presentations come from the same
user, unless this is inherent in the disclosed attributes required by their access control policies;

\item\label{c:pool} \textbf{Pooling resistance.} Colluding users cannot
make a relying party accept, in one presentation, credentials that were not
all issued to the same user;
\item\label{c:eff} \textbf{Efficiency and usability.} A presentation
consumes only the credential copies that the use case requires and
involves no party other than the holder and the relying party, so that it
also works offline, as the ARF requires for proximity presentations, which must be possible without an Internet connection~\cite[EDP\_10]{arf_annex2}; binding the credentials of a holder takes a number of attestations at most linear in the batch size and in the number of credentials to be linked together.
\end{enumerate}
As shown in Section~\ref{sec:prior}, none of the approaches of the
ARF achieves all the above properties simultaneously.

\subsection{Our Contribution}
\label{sec:contrib}

\paragraph{The
credential pooling attack.}
We have implemented a credential pooling attack on device binding in the latest version of the EUDI Wallet reference architecture
 (Figure~\ref{fig:attack}). A relying party asks Alice for two credentials, a PID and a university degree. Alice holds a PID but no degree; Bob holds
a degree. Alice and Bob collude: a proxy that they control answers the
relying party with Alice's PID and Bob's degree, and asks each of their
smartphones to sign the relying party's challenge with the key bound to its own
credential. Each signature is genuine and verifies under the key bound to
its credential, and nothing in the response relates the two keys, so the
relying party accepts the two credentials as those of one person. No key
leaves its WSCD: the smartphones only sign, and both holders must be online at
presentation time. We reproduced the attack end to end against the
unmodified reference issuer and relying party published by the European Commission~\cite{eudi_refimpl},  both with software wallets and with two Android smartphones.
The median time of the whole pooled presentation is below 0.2\,s.

The above attack would fail if the $i$-th copies of all credentials
were bound to the same key and the $i$-th presentation of the wallet spent
the $i$-th copy of every issued credential, otherwise unlinkability would be lost. This countermeasure is however concretely inapplicable for two reasons. The main one is that burning so many credentials for just one presentation can be extremely expensive.
The second is that the ARF
requires a new key pair for every credential, as specified in  Section~\ref{sec:sota}.  
\paragraph{ \proto{}: a standard-compliant countermeasure.}
We have designed a countermeasure to the credential pooling attack focusing on maintaining standardized credentials, presentations, efficiency and cryptographic tools. 
Our countermeasure, \proto{} (CREdential STApling), runs entirely within the ARF and obtains both pooling resistance and relying-party unlinkability, while preserving efficiency and usability, staying with what existing WSCDs can already compute. The key
observation is that whether two keys belong to the same holder cannot be
seen at presentation, where batch issuance makes them look unrelated, but
can be checked at issuance: a single KA generated by a Wallet Provider can attest that all the keys it
lists are in the same WSCD. This can be implemented with current Android smartphones, as shown in Section~\ref{sec:attestkey}.
An issuer can
then certify this fact in a binding attestation: a credential in
a standard format, issued to the wallet like any other, whose content
states that a set of fresh, selectively disclosable  keys was generated in the same WSCD. We
call an issuer that issues binding attestations a Binding
Attestation Issuer (BAI). This is not a new role: a BAI is an ordinary
issuer in the sense of the ARF, which issues standard credentials with
special contents, and it can be an issuer that the holder already uses, as long as it is considered trustworthy for such a task. The BAI
receives only the public keys to be bound, never the credentials or their
attributes. 
Our binding attestations therefore achieve the issuer-attested binding that the ARF lists~\cite{arf_topic_k}, and to the best of our knowledge ours is the first specified realization satisfying all properties from C1 to C6.
Figure~\ref{fig:defense} summarizes
the mechanism.

We discuss how binding attestations prevent pooling attacks exactly in the setting of the
ARF, where every single credential is bound to a key in the WSCD of its
subject and a WSCD belongs to one person. We also implemented our countermeasure:
it costs the relying party one additional signature verification (about 0.42 ms) per presentation.

\begin{figure}[!t]
\centering
\begin{tikzpicture}[
  font=\small,
  box/.style={draw, rounded corners=3pt, align=center, minimum height=1.05cm, inner sep=4pt, fill=#1},
  box/.default=white,
  arr/.style={-{Stealth[length=2.2mm]}, thick},
  lab/.style={font=\footnotesize, align=center, inner sep=1.5pt}]
\node[box=blue!6, minimum width=2.7cm] (V) at (0,0) {\textbf{Relying party}\\ asks for a PID\\ \emph{and} a degree};
\node[box=red!6, minimum width=2.4cm] (P) at (5.4,0) {\textbf{Proxy}\\ controlled by\\ Alice and Bob};
\node[box=gray!8, minimum width=3.3cm] (A) at (11.4,1.7) {\textbf{Alice's smartphone}\\ PID, key $pk_A$};
\node[box=gray!8, minimum width=3.3cm] (B) at (11.4,-1.7) {\textbf{Bob's smartphone}\\ degree, key $pk_B$};
\draw[arr] ([yshift=3mm]V.east) -- node[lab, above] {(1) request,\\ challenge} ([yshift=3mm]P.west);
\draw[arr] ([yshift=-3mm]P.west) -- node[lab, below=2pt] {(4) PID + degree,\\ two valid PoPs} ([yshift=-3mm]V.east);
\draw[arr] ([xshift=-2mm]P.north) |- node[lab, above, pos=0.7] {(2) PoP request} ([yshift=2mm]A.west);
\draw[arr] ([yshift=-2mm]A.west) -| node[lab, below, pos=0.3] {(3) signed PoP} ([xshift=2mm]P.north);
\draw[arr] ([xshift=-2mm]P.south) |- node[lab, below, pos=0.7] {(2) PoP request} ([yshift=-2mm]B.west);
\draw[arr] ([yshift=2mm]B.west) -| node[lab, above, pos=0.3] {(3) signed PoP} ([xshift=2mm]P.south);
\node[font=\footnotesize\bfseries, text=red!60!black, align=center, below=2pt of V] {(5) accepted as\\ one person};
\end{tikzpicture}
\caption{The \poolattack{}. Alice holds a PID and Bob a
university degree, each bound to a key in the secure hardware of their own
smartphone. A proxy that they control receives the relying party's request and
challenge~(1) and forwards the challenge to both smartphones as a request for a
proof of possession (PoP)~(2). Each smartphone signs it with the key bound to its
own credential and returns the signed PoP~(3), and the proxy sends both
credentials with both PoPs in one response~(4). Every signature is genuine
and no key leaves its smartphone, yet the relying party accepts the two
credentials as those of a single person~(5).}
\label{fig:attack}
\end{figure}

\section{Background}
\label{sec:part:background}
\label{sec:background}

\begin{table}[t]
\centering
\small
\begin{tabular}{@{}p{0.27\textwidth}p{0.66\textwidth}@{}}
\toprule
Term & Meaning \\
\midrule
ARF & Architecture and Reference Framework, the technical design of the EUDI Wallet \\
PID, EAA & Person Identification Data, the digital identity card; Electronic Attestation of Attributes, any other credential \\
Issuer & The organization that signs a credential \\
Relying party & The service that asks the user to present credentials \\
User, holder & The person who holds credentials in the wallet; we use the two terms interchangeably \\
Wallet Provider (WP) & The organization that provides the wallet app and vouches for its secure hardware \\
Wallet Unit & One installation of the wallet on one smartphone, with its secure hardware \\
WSCD & Wallet Secure Cryptographic Device: the smartphone's secure hardware, from which keys cannot be extracted \\
Device binding & Each credential names a key in the WSCD; presenting it requires a signature with that key \\
Key Attestation (KA) & A statement signed by the WP listing keys that were generated in one WSCD \\
Batch issuance & Issuing each credential as several single-use copies, each with its own key \\
Relying-party unlinkability & Colluding relying parties cannot tell whether two presentations come from the same user \\
Pooling & Colluding users combining credentials from different wallets in one presentation \\
Pooling resistance & A relying party can tell whether the presented credentials belong to one person \\
Attest key & An Android key that an app generates in the secure hardware; it certifies only keys generated by the same app~\cite{android_keygenspec} \\
Binding attestation & A standard credential stating that a set of keys was generated in one WSCD \\
Binding Attestation Issuer (BAI) & An ordinary issuer that issues binding attestations \\
\bottomrule
\end{tabular}
\caption{Terms used in this paper.}
\label{tab:terms}
\end{table}

\paragraph{Parties.}
Besides users, issuers and relying parties, the ARF~\cite[Section~3]{arf} defines
the Wallet Provider (\wprov{}), an organization
mandated or recognized by a Member State, which provides the wallet and vouches for
its secure hardware, and trusted list providers, which publish the
issuers, Wallet Providers and other parties that may be trusted. Issuers
are PID Providers, which issue the PID after verifying the user's identity,
and Attestation Providers, which issue EAAs.
One installation of the wallet app on a smartphone, together with its WSCD, is called a Wallet Unit.

\paragraph{Presentations.}
When a relying party requests several credentials in one presentation, the
wallet proves possession of each device key separately. In an mdoc
presentation, each document carries its own signature, computed with
that document's device key over a structure that includes the session
transcript~\cite{iso18013}. In an SD-JWT VC presentation, each credential
carries its own key-binding JWT, signed with that credential's device key
over the relying party's nonce, an audience value identifying the relying party and a
hash of the presented credential~\cite{sdjwt}. In both formats, nothing in
the response relates the device keys of different credentials to each
other.

\paragraph{Key attestation.}
An issuer learns that a key was generated by a WSCD from a Key
Attestation (KA) signed by the \wprov{}\footnote{Android also speaks of key attestation, for the certificate that the secure hardware issues for a single key; we call it Android key attestation. In this paper, KA always denotes the attestation of the ARF, signed by the \wprov{}, which can list several keys.}: a KA lists public keys and states that their private keys were generated in the WSCD~\cite{arf_annex2}. The issuer verifies the KA,
checks that it has not been revoked, and binds the credential to one of the
listed keys. Together with the Wallet Instance Attestation, which attests
the integrity of the wallet application, the KA forms the Wallet Unit
Attestation (WUA).
A relying party never sees a KA: it trusts the device binding
of a credential because the issuer checked a KA before signing the
credential. The ARF does not specify how the \wprov{} verifies that the keys
it lists were generated in the WSCD~\cite{arf_annex2}.

{\paragraph{Activation.}
\label{activation}
After the user installs the wallet application, the Wallet Unit must be
activated with its \wprov{} before any credential can be issued to it.
During activation, the \wprov{} verifies that the application is a genuine
instance of its wallet and that the device has a WSCD certified for Level
of Assurance High. It then issues to the Wallet Unit its first
KA, after verifying that the listed keys are protected by the WSCD, and its
first Wallet Instance Attestation, and re-issues both as needed during the
lifetime of the Wallet Unit~\cite{arf_annex2}.}

\paragraph{Batch issuance.}
The ARF enables every wallet to support once-only {credentials},
issued in batches of copies that are each presented only once, and
limited-time {credentials}, which are short-lived but may be
presented repeatedly~\cite{arf_annex2}. Every copy has its own device key,
and the device-binding key is among the values that must differ across
all credentials issued by that Provider. In this paper, batch issuance refers to
once-only {credentials}, and $n$ denotes the number of copies in a batch. We write $C^i_j$ for the $i$-th copy ($1\le i\le n$) of
the batch associated with the $j$-th credential ($1\le j\le k$) stored in the wallet, where $k$ is the total number of credentials held by the wallet, and
$pk^i_j$ for its device key.

\paragraph{Verifying device binding.}
The ARF states that a relying party should verify the device-binding
signature in a presentation~\cite{arf_annex2}, but no requirement
makes the check mandatory. A relying party that skips this check would accept a credential
from anyone who holds a copy, so we consider only relying parties that perform
the check.

\paragraph{Secure hardware.}
\label{sec:se}
Depending on the operating system, different features can be available to the wallet app. Here we will focus on Android smartphones.
On Android, a wallet reaches secure hardware through
the Android Keystore, which is backed either by StrongBox, implemented
in an embedded secure element or an integrated secure enclave, or by a
trusted execution environment (TEE). StrongBox supports only a few standard
algorithms, among them ECDSA and ECDH on the P-256
curve~\cite{android_keystore}. When it generates a key, the secure
hardware can also issue an attestation certificate for it, the Android key attestation, which records,
among other things, the challenge chosen by the app, whether the key was
generated or imported, and whether it lives in StrongBox or elsewhere, such as a TEE~\cite{android_attestation}. By default this certificate is signed by a
platform attestation key, either a factory key shared by many devices or a
remotely provisioned key, so it does not identify the
device~\cite{android_attestation,rkp}. 
{Starting with version 12 of Android, an app can generate its own attest key~\cite{android_keygenspec}, a key
whose only purpose is to sign the attestation certificates of other keys generated by the same app. All keys that the app generates afterward can then be attested by that attest key. The attest key can be used only for the above purpose~\cite{keymint}.}

\paragraph{Pooling adversary and trust assumptions.}
\label{sec:threat}
The specific adversary that we consider consists of a group of colluding users, each holding genuine
credentials in a genuine Wallet Unit. They may modify their wallet apps and
coordinate their action in any phase, but they cannot extract private keys from a
WSCD or forge signatures. Their goal is to violate the pooling resistance property.

As trust assumptions, we keep those of the ARF:
\begin{enumerate}[leftmargin=1.6em,itemsep=1pt,topsep=2pt]
\item the WSCD keeps private keys inside and its attestation certificates are correct;
\item the \wprov{} lists keys together in one KA only if they were generated
in the same WSCD (Section~\ref{sec:attestkey} shows how it can check this on
Android);
\item issuers certify only correct attributes, which for the Binding
Attestation Issuer of Section~\ref{sec:construction} means binding only keys
listed in the same KA;
\item a PID is issued only to a WSCD controlled by the PID's subject.
\end{enumerate}
Like the ARF, we do not protect users against issuers that collude with relying
parties. The ARF calls this Attestation Provider linkability: it cannot be fully
eliminated with credential formats based on salted hashes, and organizational
measures and audits deter it~\cite[Section~7.4.3.5]{arf}. \proto{} inherits
this limitation.
\label{sec:prior}

\paragraph{Cryptographic and issuer-attested binding.} The ARF names five approaches to provide pooling resistance \cite{arf_topic_k}, but only two of them could prevent pooling without identifying the user: cryptographic binding and issuer-attested binding. In cryptographic binding, the Wallet Unit proves to the relying party, at presentation, that the private keys of all the presented attestations were generated in the same WSCD\footnote{The ARF writes: ``Cryptographic binding between attestations is an envisioned cryptographic mechanism that enables a WSCA/WSCD or keystore to prove that it manages the private keys corresponding to two (or more) public keys''~\cite[Section~6.6.3.10.2]{arf}.}.
         The ARF does not provide the specifications to implement cryptographic binding 
         and recommends a zero-knowledge proof (ACP\_03~\cite{arf_annex2}), which relying parties would have to verify, violating \ref{c:ver}. 
          
          Verheul~\cite[Section~2.1]{verheul} proposed to relate two device keys in the
          same WSCD through a
          Proof of Association, which must be computed inside the WSCD, requiring customized instructions, therefore violating \ref{c:se}; relying parties then need to verify a
          Schnorr-type proof, violating \ref{c:ver}~\cite{verheul,arf_disc442}. 

Issuer-attested binding is described in one sentence: an issuer ``may
issue a dedicated attestation to the Wallet Unit explicitly confirming that
two or more attestations relate to the same entity''~\cite{arf_topic_k}.
If the user has to show the issuer every attestation to bind, the issuer obtains the full content of credentials that it did not issue, although it only needs to know that they belong to the same user. This violates the data minimization principle, under which personal data must be ``limited to what is necessary in relation to the purposes for which they are processed''~\cite[Article~5(1)(c)]{gdpr}, a principle that the ARF adopts~\cite[Section~4.2.4]{arf}.
On the other hand, if it is restricted to binding only attestations that it
issued itself,
it cannot bind a PID to a university degree, which is a combination explicitly mentioned in the examples in Topic~K of the discussions associated with the ARF requirements \cite{arf_topic_k}. {In both cases, a naive
unlinkable variant of the issuer-attested binding would need an attestation for each possible combination of credentials that
could be presented together, which is unfeasible since the number of binding attestations grows exponentially with the batch size and the number of credentials.
Alternatively, there could be a
request to the issuer at every presentation, but this rules out offline use
(\ref{c:eff}).}
The Spanish data protection authority judged that, realized with
privacy-preserving techniques such as zero-knowledge proofs, issuer-attested binding ``may
represent the approach most GDPR-aligned''~\cite{aepd_topick}. 

\paragraph{Other unsuitable approaches to bind attestations.}
\label{sec:sota}
\label{sec:cryptobind}
Binding all credentials of a wallet to one device key would let a relying
party check that the presented keys are equal, but that key would link
every presentation, and the ARF requires a new key pair for every PID or attestation~\cite[ISSU\_12b]{arf_annex2}.
Binding to a common key only the copies at index~$i$ of all batches of
credentials would avoid this linking, but it would be extremely expensive,
since a presentation would waste the copies at its index of credentials it
does not even refer to. 

\section{Our Proposal: \proto{}}
\label{sec:part:linker}
\label{sec:idea}

In order to prevent pooling attacks, someone must check that the keys of the presented
credentials have been generated by  the same WSCD. The relying party cannot run this check: batch
issuance makes the keys of one user look uncorrelated.
At issuance, however,
the check is possible, because {a single KA, signed by the \wprov{}, can list
several keys and state that they were all generated in the same WSCD.}  A
binding attestation carries this fact from issuance, where it can
be checked, to presentation, where it is needed.

\paragraph{Attest keys in Android.}
\label{sec:attestkey}
At activation, the wallet generates an attest key $ak$ in its WSCD, and the
\wprov{} verifies the attestation certificate of $ak$, signed by the platform attestation key (Section~\ref{sec:se}), and records $ak$ for that
Wallet Unit. Every device key that the wallet later generates is attested
under $ak$. The WSCD uses $ak$ only to attest keys that it generates, so two
keys attested under the same $ak$ have been generated in the same WSCD, and no modified
wallet can have Alice's $ak$ attest a key that was generated in the WSCD of Bob's smartphone. The
\wprov{} therefore lists keys in one KA only if all of them are attested
under the $ak$ it recorded for the Wallet Unit, and were generated, not
imported, in the expected secure hardware. A KA listing several keys is
then sound evidence that they are within the same WSCD, and producing it needs nothing
beyond what is available in many existing Android smartphones. The attest key is a stable
identifier of the Wallet Unit, but only the \wprov{}, which already knows
the Wallet Unit, ever sees it: issuers see KAs, and relying parties see neither.

\subsection{How \proto\ Works}
\label{sec:construction}

We now describe how our countermeasure \proto\ works within the ARF.

Consider a holder who wants to present together two credentials, $C_1$
from issuer $A$ and $C_2$ from issuer $B$, each issued in a batch of $n$
copies. For simplicity we describe the case of two credentials; in Section~\ref{sec:when} we generalize it to $k$ credentials. An issuer acts as a Binding Attestation Issuer ($\linker$)
when it verifies a KA signed by a \wprov{} on a trusted list and issues, in
a standard credential format, a binding attestation whose content is a set
of keys listed in that KA. Any issuer recognized through a trusted list can
act as a $\linker$. We distinguish four phases:
\begin{enumerate}[leftmargin=1.6em,itemsep=1pt,topsep=2pt]
    \item Activation: this procedure follows the one of the ARF (see Section~\ref{activation}), and in addition the wallet
    generates an attest key $ak$ in its WSCD,
    which the \wprov{} verifies and records for the Wallet Unit;
    \item Paired key generation: the wallet generates $n$ key
    pairs {$(pk^i_1,pk^i_2)$}, for all ($1\le i\le n$), in its WSCD, each attested under $ak$;
    \item Issuance: the user runs two ordinary OpenID4VCI procedures with
    issuer A and issuer B and obtains two batches of  $n$ copies each of credentials
    $C^i_1$ and $C^i_2$, bound to the keys $pk^i_1$ and $pk^i_2$ for all  ($1\le i\le n$), respectively;
    \item Binding attestation issuance: the wallet obtains from the
    \wprov{} a KA listing the keys {$pk^i_1$ and $pk^i_2$} for all ($1\le i\le n$) 
    and sends it to $\linker$ only. $\linker$ verifies
    the KA and issues, for all ($1\le i\le n$), {the binding attestation
    $\link^i$}, 
{an SD-JWT VC signed by $\linker$ whose selectively disclosable attributes are the keys $(\{pk^i_1,pk^i_2\})$.} 
\end{enumerate}
Notice that $\linker$ receives only the public keys listed in the KA. It
never receives $C_1$ or $C_2$, nor any of their attributes. 

\paragraph{Presentation.}
The wallet selects an unused index $i \in \{1,\ldots,n\}$, sends $C^i_1$, $C^i_2$ and
$\link^i$ to the relying party, signs the session transcript with both keys, and {never uses $C^i_1$, $C^i_2$ or $\link^i$ in other presentations}. {Below, $pk_1$ and $pk_2$ denote the keys $pk^i_1$ and $pk^i_2$ of the presented copies.}

\paragraph{What the relying party checks.}
The relying party performs three checks:
\begin{enumerate}[label=\arabic*.,leftmargin=1.6em,itemsep=1pt,topsep=2pt]
\item run the standard OpenID4VP verification of a combined credentials presentation, extracting $pk_1$ and $pk_2$ from $C_1$ and $C_2$ respectively;
\item verify $\linker$'s signature on $\link$ against the trusted list;
\item check that the keys certified in $\link$ are exactly $\{pk_1,pk_2\}$.
\end{enumerate}
Step 1 is what a relying party already does for a combined
presentation; steps~2 and~3 add one signature verification, that of $\linker$ on $\link$, and an equality
comparison, so the relying party still checks only ordinary signatures.

\begin{figure}[!t]
\centering
\begin{tikzpicture}[
  font=\small,
  box/.style={draw, rounded corners=3pt, align=center, minimum height=1.25cm, inner sep=4pt, fill=#1},
  box/.default=white,
  arr/.style={-{Stealth[length=2.2mm]}, thick},
  lab/.style={font=\footnotesize, align=center, inner sep=1.5pt},
  phase/.style={font=\footnotesize\itshape, text=gray!70!black, anchor=west}]
\node[phase] at (-1.7,2.05) {At issuance};
\node[box=gray!8, minimum width=3.3cm] (W) at (0,0) {\textbf{Holder's smartphone}\\ its attest key $ak$\\ certifies $pk_1$ and $pk_2$};
\node[box=blue!6, minimum width=3.3cm] (WP) at (5.9,0) {\textbf{Wallet Provider}\\ both under the same $ak$\\ $\Rightarrow$ one KA lists both};
\node[box=green!8, minimum width=3.3cm] (I) at (11.6,0) {\textbf{BAI} (an issuer)\\ checks the KA, issues\\ $\link\mathsf{\ on}\ pk_1\ and\ pk_2$};
\draw[arr] (W) -- node[lab, above=2pt] {(1) certificates\\ of $pk_1,pk_2$} (WP);
\draw[arr] (WP) -- node[lab, above=2pt] {(2) KA} (I);
\draw[arr] (I.north) to[bend right=14] node[lab, above] {(3) $\link$, stored by the wallet} (W.north);
\node[phase] at (0.15,-1.5) {At presentation};
\node[box=white, minimum width=3.3cm] (V) at (11.6,-2.6) {\textbf{Relying party}\\ $\link$ covers $pk_1,pk_2$\\ $\Rightarrow$ one person};
\draw[arr] (W.south) |- node[lab, above, pos=0.72] {(4) $C_1$ (key $pk_1$), $C_2$ (key $pk_2$), $\link$, two PoPs} (V.west);
\node[font=\footnotesize, text=red!60!black, align=center] at (5.8,-3.75)
 {\textbf{Pooling fails:} Alice's $ak$ cannot certify Bob's key $\Rightarrow$ no KA lists both
 $\Rightarrow$ no $\link$ $\Rightarrow$ rejected};
\end{tikzpicture}
\caption{\proto{}, our countermeasure, for two credentials $C_1$
and $C_2$. At issuance, the attest key $ak$ in the holder's smartphone certifies
the device keys $pk_1$ and $pk_2$ of the two credentials~(1). The Wallet
Provider lists both keys in one Key Attestation (KA) only if the same $ak$
certified them~(2), and the Binding Attestation Issuer (BAI), an ordinary
issuer, checks the KA and returns a binding attestation $\link$ on
$\{pk_1,pk_2\}$, a credential in a standard format~(3). At presentation, the
wallet sends $\link$ with the two credentials~(4), and the relying party
checks that $\link$ covers their keys. Alice and Bob cannot obtain a $\link$
on their two keys, because no attest key certifies keys in two smartphones.}
\label{fig:defense}
\end{figure}

\paragraph{More than two credentials.}
Section~\ref{sec:when} shows how a single binding attestation per batch element
covers any combination of the $k$ credentials, avoiding the exponential number of
attestations a naive scheme would need.

\subsection{When Binding Attestations Are Issued}
\label{sec:when}

The
wallet can obtain its binding attestations
right after obtaining some credentials, for instance from their issuer. Alternatively, the wallet can obtain a binding attestation when a presentation with multiple credentials needs to start. In this last case, the wallet sends the public keys of the requested credentials to $\linker$, receives a binding attestation over
them, and forwards it to the relying party. In both cases $\linker$ checks
a KA, but
each option has advantages and disadvantages.

\paragraph{BA at presentation.}
Obtaining binding attestations at presentation has two shortcomings.
\begin{itemize}[leftmargin=1.2em,itemsep=1pt,topsep=2pt]

\item \emph{Latency.} Every combined presentation waits for additional
online round trips to the \wprov{} and to $\linker$, within the relying
party's session, and depends on the network and on the availability of
both.
\item \emph{Connectivity.} A holder without connectivity cannot obtain a
binding attestation, and so cannot complete any combined presentation, even
in proximity flows, which ISO/IEC 18013-5 supports without an Internet
connection~\cite{iso18013}.
\end{itemize}

\paragraph{BA at issuance.}
Obtaining binding attestations at issuance circumvents the above shortcomings: the
\wprov{} and $\linker$ are contacted only outside presentations, when the
wallet obtains credentials or binding attestations, they
learn nothing about individual presentations, and presentations proceed
offline with no additional round trip. At issuance, however, the wallet
does not know which credentials a relying party will later request
together. If a binding attestation states exactly the set of keys to be
presented, covering every combination of $k$ credentials needs one binding
attestation for each subset of them, which is clearly infeasible.

\begin{figure}[t]
\normalcolor
\centering
\begin{tikzpicture}[
  font=\footnotesize,
  frame/.style={draw, rounded corners=3pt, fill=blue!5},
  dig/.style={draw, fill=white, minimum width=2.1cm, minimum height=0.45cm, inner sep=1pt},
  hid/.style={dig, fill=gray!15, text=gray!60!black},
  disc/.style={draw, rounded corners=2pt, fill=green!8, minimum width=3.3cm, align=center, inner sep=2pt},
  off/.style={draw=gray!60, dashed, rounded corners=2pt, minimum width=3.3cm, align=center, inner sep=2pt, text=gray!60!black},
  arr/.style={-{Stealth[length=1.8mm]}},
  note/.style={align=left, text width=7.0cm, anchor=north west, inner sep=0pt}]
\node[anchor=west, font=\small\bfseries] at (0,0.45) {(a) Issued by $\linker$, stored by the wallet};
\draw[frame] (0,0) rectangle (3.2,-3.95);
\node[font=\footnotesize\bfseries] at (1.6,-0.3) {Issuer-signed JWT};
\node at (1.6,-0.68) {signed by $\linker$};
\node[rotate=90] at (0.45,-2.35) {\texttt{keys}};
\node[dig] (a1) at (1.95,-1.25) {$H(d^i_2)$};
\node[dig] (a3) at (1.95,-2.35) {$H(d^i_1)$};
\node[dig] (a5) at (1.95,-3.45) {$H(d^i_3)$};
\node[disc] (ad2) at (5.6,-1.25) {$d^i_2=(r^i_2,\,pk^i_2)$\\ {\scriptsize key of the degree}};
\node[disc] (ad1) at (5.6,-2.35) {$d^i_1=(r^i_1,\,pk^i_1)$\\ {\scriptsize key of the PID}};
\node[disc] (ad3) at (5.6,-3.45) {$d^i_3=(r^i_3,\,pk^i_3)$\\ {\scriptsize key of the driving license}};
\draw[arr] (ad2.west) -- node[above]{$H$} (a1.east);
\draw[arr] (ad1.west) -- node[above]{$H$} (a3.east);
\draw[arr] (ad3.west) -- node[above]{$H$} (a5.east);
\node[note] at (0,-4.25) {$\linker$ signs only salted digests: no key appears in the clear. Each digest is a combination of a public key 
and a random value.};
\begin{scope}[xshift=8.4cm]
\node[anchor=west, font=\small\bfseries] at (0,0.45) {(b) Presented for a PID and a degree};
\draw[frame] (0,0) rectangle (3.2,-3.95);
\node[font=\footnotesize\bfseries] at (1.6,-0.3) {Issuer-signed JWT};
\node at (1.6,-0.68) {unchanged};
\node[rotate=90] at (0.45,-2.35) {\texttt{keys}};
\node[dig, very thick] (b1) at (1.95,-1.25) {$H(d^i_2)$};
\node[dig, very thick] (b3) at (1.95,-2.35) {$H(d^i_1)$};
\node[hid] (b5) at (1.95,-3.45) {$H(d^i_3)$};
\node[disc] (bd2) at (5.6,-1.25) {$d^i_2$ sent\\ {\scriptsize $pk^i_2$, key of $C^i_2$}};
\node[disc] (bd1) at (5.6,-2.35) {$d^i_1$ sent\\ {\scriptsize $pk^i_1$, key of $C^i_1$}};
\node[off] (bd3) at (5.6,-3.45) {$d^i_3$ not sent\\ {\scriptsize the license key stays hidden}};
\draw[arr] (bd2.west) -- node[above]{$H$} (b1.east);
\draw[arr] (bd1.west) -- node[above]{$H$} (b3.east);
\node[note] at (0,-4.25) {The relying party verifies the signature of
$\linker$ once, finds $H(d^i_1)$ and $H(d^i_2)$ in \texttt{keys}, and checks
that $pk^i_1$ and $pk^i_2$ are the device keys of the presented PID $C^i_1$
and degree $C^i_2$.};
\end{scope}
\end{tikzpicture}
\caption{Structure of the binding attestation $\link^i$
of a holder with three credentials: a PID $C_1$, a degree $C_2$ and a
driving license $C_3$. (a)~What $\linker$ issues and the wallet stores.
(b)~What the wallet presents to a relying party that requests a PID and a
degree.}
\label{fig:ba}
\end{figure}

\paragraph{Removing the trade-off: the best of both worlds.}
Selective disclosure in SD-JWT VCs lets us bind at issuance and choose the
combination at presentation: a binding attestation lists the keys as
selectively disclosable attributes. Let $pk^i_1,\dots,pk^i_k$, $i \in \{1,...,n\},$ be the device
keys of the $i$-th copies of the holder's $k$ credentials. 
For each  ($1\le i\le n$), $\linker$ issues a single binding attestation
$\link^i$ over attributes
$H(d^i_1),\dots,H(d^i_k)$,
where $d^i_j=(r^i_j,\,pk^i_j)$ for $1\leq j \leq k$, $H$ is the cryptographic  hash function of the SD-JWT and each $r^i_j$ is a fresh
random salt. 
The holder stores $\link^i$ with the
disclosures $d^i_1,\dots,d^i_k$ (Figure~\ref{fig:ba}). When the relying party requires attributes belonging to 
$\ell$ credentials stored in the wallet, the wallet selects an index~$i \in \{1,...,n\}$ 
such that $\link^i$ and the $i$-th copies of those $\ell$ credentials are all unused, and sends these copies together with $\link^i$, disclosing the requested attributes of the credentials and, in $\link^i$, only the $\ell$ corresponding keys. The
relying party verifies $\link^i$ like any SD-JWT VC, with a single
signature verification, and performs the checks of
Section~\ref{sec:construction} on the disclosed keys.

The combination is thus chosen at presentation, through the disclosures,
and yet no party is contacted at presentation: a single binding
attestation, issued in advance, covers all  combinations at once,
so the holder needs $n$ binding attestations, one per batch element, and the non-disclosed keys stay
hidden  and thus can still be disclosed individually without any linkability issue. 

\paragraph{Remark on hiding the number of credentials.} 
A binding attestation includes as (selectively disclosable) attributes keys belonging to credentials of the user. Since a standard presentation does not hide the number of included attributes, the relying party learns the number of credentials that the wallet decided to link through a binding attestation. In applications in which this is considered a privacy leak, the wallet can simply ask the binding attestation issuer to add several random digests as attributes and to shuffle all attributes, so that the leak is then practically canceled, since only a large  upper bound on the number of credentials is revealed.

\subsection{Why Binding Attestations Prevent Pooling}
\label{sec:analysis}

A relying party accepts a combined presentation only if a binding attestation
from a trusted issuer covers exactly the device keys of the presented
credentials, and the presentation carries a valid signature under each of
those keys. Tracing a binding attestation back to its origin shows why
colluding holders cannot obtain one for credentials of different people:
\begin{itemize}[leftmargin=1.2em,itemsep=1pt,topsep=2pt]
\item the Binding Attestation Issuer certifies only keys that are listed
together in one KA;
\item the \wprov{} lists keys together in one KA only if all of them are
attested under the attest key that it recorded for one Wallet Unit;
\item the secure hardware uses an attest key only to attest key pairs stored in that same secure hardware, and marks each as \texttt{GENERATED} or \texttt{IMPORTED}, so the WP can require keys generated inside the secure hardware. 
\end{itemize}
The keys covered by a binding attestation therefore were generated in the same
WSCD. In the setting that the ARF assumes, every credential is bound to a
key in the WSCD of its subject, and a WSCD belongs to one person, so all
the credentials covered by one binding attestation have the same subject.
In the credential pooling attack, Alice's PID key is generated in her smartphone and Bob's degree
key in his: no attest key attests both, so no KA lists both, and no issuer
certifies them together. Nor can the colluders forge a binding
attestation, a KA or a device signature, since all of them are ordinary
ECDSA signatures.
The argument above relies only on the trust assumptions of the threat model (Section~\ref{sec:threat}).

\paragraph{Limitations.}
A binding attestation states that the keys belong to the same WSCD at issuance. That this still holds at presentation follows from the secure hardware, since
no key can leave its WSCD. The mechanism also needs a way for the
\wprov{} to tell that keys belong to  a WSCD, and this leverages  attest keys on Android.
\proto{} follows a general-purpose design and the example we have given for existing versions of Android can be replaced by other mechanisms to include other operating systems (e.g., iOS) and future versions of Android.

\section{Implementation and Evaluation}
\label{sec:part:impleval}
\label{sec:impl}
\label{sec:eval}

\paragraph{Testbed.}
We tested both the attack and the countermeasure against the official  reference issuer (\nolinkurl{eudi-srv-pid-issuer}, commit
\texttt{3ebd78f}) and reference relying-party service
(\nolinkurl{eudi-srv-web-verifier-endpoint-23220-4-kt}, commit
\texttt{d74c971}), both unmodified and run from their published container
images. Our wallet and pooling proxy are built on the official EUDI
libraries (\texttt{openid4vp-kt} 0.15.1, \texttt{openid4vci-kt} 0.13.1,
\texttt{sdjwt-kt} 0.20.1) and use the deployed OpenID4VP~1.0 and SD-JWT~VC
profile. The two colluding holders run either as software processes or on
two Android smartphones: a Samsung Galaxy~S24~Ultra, with its key in StrongBox,
and a Galaxy~A53, with its key in the TEE. On each smartphone, a small app
generates the device key in the Android Keystore and signs whatever the
proxy sends. Software measurements run on a desktop with an AMD
Ryzen~9, OpenJDK~21 and Linux. All experiments use our own devices
and sample credentials.

\paragraph{The pooling attack.}
\label{sec:eval:repro}
We reproduced the scenario of Figure~\ref{fig:attack}. Alice's smartphone holds
a PID, and Bob's a university degree issued to a different person. The relying party sends a single request for Alice's
birthdate from a PID and for a graduation date from a degree. The proxy
answers it, asks each smartphone to sign the relying party's challenge with the key
bound to its credential, and returns both credentials in one response.
The reference relying party, which checks every device signature, accepts the
response as coming from a single person. This holds both with software
holders and with the two smartphones.

\paragraph{Latency.}
\label{sec:eval:q1lat}
Table~\ref{tab:attackA} reports the end-to-end time of the pooled
presentation. Since the proxy asks for the two signatures in parallel, the
pooled presentation over two smartphones (\num{147}\,ms over USB, \num{182}\,ms
over WiFi) takes about as long as an honest presentation signed in
StrongBox (\num{139}\,ms). Timing therefore cannot tell a pool from an
honest holder, and all these  latencies are far below realistic session timeouts:
the reference relying party defaults to \num{6400}\,min.

\begin{table}[h]
\centering
\small
\begin{tabular}{@{}lr@{}}
\toprule
Setup & Median (ms) \\
\midrule
\multicolumn{2}{@{}l}{\emph{Pooled presentation (two credentials, two holders)}}\\
\quad software holders, one host            & 75 \\
\quad two smartphones, USB, sequential signing   & 166 \\
\quad two smartphones, USB, parallel signing     & 147 \\
\quad two smartphones, WiFi, parallel signing    & 182 \\
\midrule
\multicolumn{2}{@{}l}{\emph{Honest presentation, one credential (USB)}}\\
\quad PID on Galaxy S24 Ultra (StrongBox)   & 139 \\
\quad degree on Galaxy A53 (TEE)            & 91 \\
\bottomrule
\end{tabular}
\caption{End-to-end latency of the \poolattack{} on the remote OpenID4VP
flow, median of \num{25} runs.}
\label{tab:attackA}
\end{table}

\paragraph{\proto{}.}
\label{sec:eval:defend}
We implemented CRESTA, our fix, by adding a Binding Attestation Issuer, which runs as a separate issuer with its own key pair, and by adding to the unmodified reference implementation's relying party an external check performing steps~2 and~3 of Section~\ref{sec:construction}. The reference implementation's issuer is unchanged.

In the testbed, the WSCDs and the attestation PKI are emulated on the model of KeyMint, the Android interface to the secure hardware that generates and attests keys~\cite{keymint}: each key generated in an emulated WSCD comes with its Android key attestation, signed by the attest key of its WSCD, whose own certificate is signed by an emulated platform attestation key (Section~\ref{sec:se}).

In deployment, the \wprov{} checks the Android key attestation for each key generated in the WSCD
and, if one attest key certifies all the keys, the \wprov{} generates a KA that lists them (Section~\ref{sec:attestkey}); the $\linker$ then verifies this KA.
For concreteness, in the testbed the $\linker$ runs the check of the \wprov{} itself
and binds the keys only if one attest key certifies them all. A genuine presentation of two credentials whose keys were generated by the same WSCD is accepted. The pooled response is now rejected, since it includes two keys belonging to two different Wallet Units and no $\link$ binds  them. Our experiments show that ARF relying parties
need no change other than an additional condition on the access control policy to resist pooling attacks. 
On Android, the attest keys of Section~\ref{sec:part:linker} ensure that colluding
holders cannot obtain a valid $\link$ for keys in two different devices.
The
testbed also simplifies the construction in two ways: binding attestations
are issued on demand at presentation rather than in advance, and they are
plain signed objects rather than SD-JWT VCs.

\paragraph{Cost.}
\label{sec:eval:q3cost}
Table~\ref{tab:cost} reports the overhead of the countermeasure. The binding check adds to the relying party one signature verification, which takes \num{0.42}\,ms, and a set comparison; the presentation grows by the
binding attestation (\num{1342}\,B).

\begin{table}[h]
\centering
\small
\begin{tabular}{@{}lcc@{}}
\toprule
Metric & Baseline & With $\link$ \\
\midrule
Relying-party CPU (ms)                 & 0.84  & 1.27 \\
Presentation size (B)             & 6755 & 8097 \\
\bottomrule
\end{tabular}
\caption[Overhead of the countermeasure]{Overhead of the countermeasure, measured on the desktop with P-256 ECDSA
(JCA \texttt{SHA256withECDSA}). Baseline: the verification of an ordinary combined presentation of two credentials; with $\link$: the same plus the binding check. Medians of \num{4001} runs.}
\label{tab:cost}
\end{table}

\FloatBarrier
\section{Concluding Remarks}
\label{sec:conclusion}
\label{sec:part:discussion}
\label{sec:discussion}

Device binding in the EUDI Wallet establishes that a single credential is being presented with the participation of the device it was issued to. It does not establish that the credentials
presented together belong to one person, and batch issuance, introduced
for privacy, removes the signal a relying party would need to check it. The ARF
foresees a cryptographic binding of attestations, but specifies no
mechanism for it. The other approaches that the ARF retains either
fail against pooling or identify the user
 (Section~\ref{sec:prior}). We have
shown that the problem is exploitable today, against the reference
implementation and with keys held in the secure hardware of commercial
smartphones. We have also shown a countermeasure, \proto{}, that stays within the ARF:
binding attestations are standard credentials, issued by ordinary issuers
and checked by relying parties like any other signature, and they rely only on
capabilities that are available in a large number of currently available smartphones. The price is that the
issuer of binding attestations
must be trusted to certify honestly which keys share a WSCD.

\paragraph{Recommendations.}
Our findings suggest two changes to the ARF. First, holder collusion
should be more explicitly part of the threat model for combined presentations.
Second, 
relying parties that request several credentials together should require a binding attestation covering their keys, as in \proto{}.

\label{sec:disclosure}

\paragraph{Ethical considerations.}
The credential pooling attack concerns a system in pre-deployment. All experiments ran
against the open-source reference implementations, deployed on our own
machines. No production issuance or verification service was contacted, no real PID was
obtained, and no third party's credential was used.
The two holders in
every experiment are researchers involved in this work, consenting to the
collusion, so no participant is deceived and no
personal data of any non-author is processed.

\subsection*{Acknowledgments}
We thank Dario Castellano and Pierpaolo Della Monica for
helpful discussions on  credential pooling attacks.

\bibliographystyle{unsrturl}
\bibliography{reference}

\end{document}